# Sedenion algebra for three lepton/quark generations and its relations to SU(5)

Qiang Tang[1] and Jau Tang[2*]

[1]School of Artificial Intelligence, Anhui University of Science and Technology, Huainan, Anhui 232000, China

[2]Wuhan University, Wuhan 430072, China

Corresponding author: Jau Tang: jautang@whu.edu.cn

**Abstract**

In this work, we analyze two models beyond the Standard Model's descriptions that make ad hoc hypotheses of three point-like lepton and quark generations without explanations of their physical origins. Instead of using the same Dirac equation involving four anti-commutative matrices for all such structure-less elementary particles, we consider in the first model the use of sixteen direct-product matrices of quaternions that are related to Dirac's gamma matrices. This associative direct-product matrix model could not generate three fermion generations satisfying Einstein's' mass-energy relation. We show that sedenion algebra contains five distinct quaternion sub-algebras and three octonion sub-algebras but with a common intersecting quaternion algebra. This model naturally leads to precisely three generations as each of the non-associative octonion sub-algebra leads to one fermion generation. Moreover, we demonstrate the use of basis sedenion operators to construct twenty-four 5x5 generalized lambda matrices representing SU(5) generators, in analogy to the use of octonion basis operators to construct Gell-Mann's eight 3x3 lambda matrices for SU(3) generators. Thus, this work links the sedenion algebra to Georgi-Glashaw's SU(5) Grand Unification Theory (GUT) that unifies the electroweak and strong interactions for known Standard Model's elementary particles.

**Keywords:** Standard Model, SU(5) GUT, sedenion, octonion, quarternion, direct-product matrices

## 1. Introduction

The Dirac equation is the Standard Model's cornerstone for describing all fermionic elementary particles [1-3]. According to the ad hoc assumptions, three generations of leptons and quarks are point-like objects without a physical size [4-6]. However, such a long-held conceptual picture of a point-like particle is inconsistent with observations. For example, why there exist precisely three generations of leptons and quarks [7], and why could a point-like muon or tau be much heavier than an electron [8, 9]. And, for the six types of point-like quarks, why are charm, strange, top, and bottom quarks heavier than up and down quarks [10, 11]. In Dirac's theory of a relativistic electron, he coupled four 4x4 anti-commutative matrices $\{\alpha_1, \alpha_2, \alpha_3, \beta\}$, or equivalently $\{\gamma^0, \gamma^1, \gamma^2, \gamma^3\}$, to the 1st-order derivatives in space and time to describe the wave [12, 13]. By extending the operator techniques beyond Dirac's theory, one can include additional operators in a higher dimension to describe a particle's internal degrees of freedom. In this work, we analyze two types of higher dimensional generalization using the direct-product operators constructed from Pauli's matrices versus hyper-complex operators beyond quaternions [14-15] which is equivalent to Dirac's 4x4 gamma-matrices. Like the 4-element quaternion algebra with three anti-commutative basis operators [16], the 8-element octonion algebra contains seven anti-commutative basis operators, but unlike the associative quaternion algebra, the octonion algebra is non-associative [17-20]. In this work, we will examine the corresponding multiplication rules and compare their multiplication tables to clarify the similarities, differences, and the physical implications.

In Sec. 2.1 we shall briefly review Dirac's original model and its connection to the quaternion algebra. In Sec. 2.2, we consider a higher dimensional model beyond Dirac's use of four anti-commutative gamma matrices. We analyze the group of 16 direct-product operators constructed from quaternions or, equivalently, Pauli's matrices and an identity operator, which are related to 16 Dirac's gamma. In Sec. 2.3, we discuss the non-associative 16-element sedenion algebra [21-24], and show it is composed of three distinctive octonion algebras. We shall analyze its differences from the direct-product case, and examine whether the corresponding mass-energy relation is consistent with Einstein's special relativity. In Sec. 2.5, we shall show how proper operator assignment of the operators from three octonion sub-algebras can naturally lead to the rise of three generations of lepton and quarks. Finally, in Sec. 2.6, we shall present a mapping of octonions to 8 SU(3) generators, and a mapping of sedenion operators to 24 generators for SU(5), which was proposed by Geogi-Glashow [25,26] for their GUT model for grand unification of the Standard Model's elementary particles.

1. **Theory**

In this section, we present a theoretical analysis of an associative algebra model using Dirac's gamma matrices, which are related to direct-product operators which are constructed from quaternions or equivalently Pauli's matrices and an identity matrix, versus non-associative algebra model based on octonion and sedenion operators. This work extends Dirac's original theory for the electron to three generations of leptons and quarks, and extends the Standard Model beyond the ad hoc assumption of a point-like elementary fermions. We shall show that the sedenion algebra provides a key link to SU(5) symmetry of GUT. This hyper-complex operator model with higher degrees of freedom leads precisely to three generations of leptons/quarks which possess internal structures.

**2.1. Dirac equation**

According to Dirac's theory for the electron, using the gamma matrices one has

$$(i\gamma^\mu \partial_\mu - m)\Psi = 0, \quad \mu = 0,1,2,3$$
$$p_k = -i\partial_k, \quad p_0 = i\partial_0 \quad \quad (1A)$$
$$\gamma^k = = \sigma_t \otimes \sigma_k = \begin{pmatrix} 0 & \sigma_k \\ -\sigma_k & 0 \end{pmatrix}, \quad \gamma^0 = = \sigma_3 \otimes \sigma_0 = \begin{pmatrix} \mathbf{I}_2 & 0 \\ 0 & -\mathbf{I}_2 \end{pmatrix}.$$

where the natural unit $\hbar = c = 1$ is used. Here, we define five 2x2 matrices, including Pauli's matrices $\sigma_1$, $\sigma_2$, $\sigma_3$, , an identity matrix $\mathbf{I}_2$ and $\sigma_t \equiv i\sigma_2$ , as

$$\sigma_1 = \begin{pmatrix} 0 & 1 \\ 1 & 0 \end{pmatrix}, \quad \sigma_2 = \begin{pmatrix} 0 & -i \\ i & 0 \end{pmatrix}, \quad \sigma_3 = \begin{pmatrix} 1 & 0 \\ 0 & -1 \end{pmatrix}, \quad \sigma_0 = \mathbf{I}_2 = \begin{pmatrix} 1 & 0 \\ 0 & 1 \end{pmatrix}, \quad (1B)$$

Equivalently, using the matrices $\alpha_k = \gamma^0 \gamma^k$ and $\beta = \gamma^0$, the Dirac equation can also be expressed as $E\Psi = (\boldsymbol{\alpha} \cdot \mathbf{P} + \boldsymbol{\beta} m)\Psi$. The identity matrix and three Pauli matrices form the basis operators for quaternion algebra. Based on Dirac's 1st-order differential equation in spacetime one obtains Einstein's relativistic mass-energy relation $E^2 = m^2 + p^2$. These Dirac's gamma matrices in $\{I_4, \gamma^0, \gamma^1, \gamma^2, \gamma^3\}$ are used in the Dirac equation. However, the product of these matrices does not satisfy the closure property, for example, $\gamma^1\gamma^2 = \sigma_0 \otimes \sigma_3$ does not belong to the same set, therefore, these five operators do not form a group.

In the Standard Model the same Dirac equation is used for all leptons and quarks, assuming these particles are a point-like object with an infinitely small volume, it does not offer physical explanations to exist precisely why there are three generations. To generalize Dirac equation to higher dimensions, and to account for three fermion generations by incorporating internal structural dynamics, we consider in the following sections two modelling approaches, i.e., an direct-product matrix model of 16 associative 4x4 matrices versus non-associative hype-complex algebra of 16 sedenion basis operators..

## 2.2. Associative algebra of 16 direct-product matrices
$\{I, \Gamma_1, \Gamma_2, \Gamma_3, \Theta_1, U_1, U_2, U_3, \Theta_2, V_1, V_2, V_3, \Theta_3, W_1, W_2, W_3\}$

To generalize Dirac's description of the electron using four Dirac's gamma matrices $\gamma^\mu (\mu=0,1,2,3)$, we can define a set of direct-product operators using three Pauli's matrices and an identity matrix as

$$\sigma_{ij} \equiv \sigma_i \otimes \sigma_j$$
$$U_k = i\sigma_{1k}, \quad V_k = i\sigma_{2k} \quad W_k = i\sigma_{3k}, \Gamma_k = i\sigma_{0k} \quad \Theta_k = i\sigma_{k0} \tag{2}$$

One can show that the eight-element set of direct-product operators $\{I, \Gamma_1, \Gamma_2, \Gamma_3, \Theta_1, U_1, U_2, U_3\}$ form a group with the closure property for multiplication. Similarly, one can show that the other two sets $\{I, \Gamma_1, \Gamma_2, \Gamma_3, \Theta_2, V_1, V_2, V_3\}$ and $\{I, \Gamma_1, \Gamma_2, \Gamma_3, \Theta_3, W_1, W_2, W_3\}$ form a group for multiplication. These direct-product operators can be related to the sixteen direct-product matrices of quaternions. The multiplication table for the sixteen direct-product operators, which are related to Dirac's gamma matrices, is illustrated in Fig. 1.

[Figure 1 multiplication table image]

**Figure 1**. The color-coded multiplication table for 16 direct-product operators. Among the sixteen operators, the 4-th, 8-th and 12-th operators are commutative with other operators although most are anti-commutative. These properties differs from that of sedenion algebra. This 16-element group contains three 8-element sub-groups with closure property for multiplication, namely, $\{\mathbf{I}, \mathbf{\Gamma}_1, \mathbf{\Gamma}_2, \mathbf{\Gamma}_3, \mathbf{\Theta}_1, \mathbf{U}_1, \mathbf{U}_2, \mathbf{U}_3\}$, $\{\mathbf{I}, \mathbf{\Gamma}_1, \mathbf{\Gamma}_2, \mathbf{\Gamma}_3, \mathbf{\Theta}_2, \mathbf{V}_1, \mathbf{V}_2, \mathbf{V}_3\}$ and $\{\mathbf{I}, \mathbf{\Gamma}_1, \mathbf{\Gamma}_2, \mathbf{\Gamma}_3, \mathbf{\Theta}_3, \mathbf{W}_1, \mathbf{W}_2, \mathbf{W}_3\}$. Each domain containing quaternion triplets follows cyclic multiplication rules.

Here we summarize their multiplication rules for these sixteen direct-product operators

$$\begin{aligned}
&\mathbf{U}_i \mathbf{U}_j = -\varepsilon_{ijk}\mathbf{\Gamma}_k - \delta_{ij}\mathbf{I}, \; \mathbf{V}_i \mathbf{V}_j = -\varepsilon_{ijk}\mathbf{\Gamma}_k - \delta_{ij}\mathbf{I}, \; \mathbf{W}_i \mathbf{W}_j = -\varepsilon_{ijk}\mathbf{\Gamma}_k - \delta_{ij}\mathbf{I} \\
&\mathbf{\Gamma}_i \mathbf{\Gamma}_j = -\varepsilon_{ijk}\mathbf{\Gamma}_k - \delta_{ij}\mathbf{I}, \quad \mathbf{\Theta}_i \mathbf{\Theta}_j = -\varepsilon_{ijk}\mathbf{\Theta}_k - \delta_{ij}\mathbf{I} \\
&\mathbf{U}_i \mathbf{V}_j = -i\varepsilon_{ijk}\mathbf{W}_k - \delta_{ij}\mathbf{\Theta}_3, \\
&\mathbf{V}_j \mathbf{U}_i = -i\varepsilon_{ijk}\mathbf{W}_k + \delta_{ij}\mathbf{\Theta}_3 \\
&\mathbf{V}_i \mathbf{W}_j = -i\varepsilon_{ijk}\mathbf{U}_k + \delta_{ij}\mathbf{\Theta}_1, \\
&\mathbf{W}_j \mathbf{V}_i = -i\varepsilon_{ijk}\mathbf{U}_k + \delta_{ij}\mathbf{\Theta}_1, \\
&\mathbf{W}_i \mathbf{U}_j = -i\varepsilon_{ijk}\mathbf{V}_k + \delta_{ij}\mathbf{\Theta}_2
\end{aligned} \quad (3A)$$

and

$$\mathbf{U}_i\mathbf{\Gamma}_j = -\sigma_{1i}\sigma_{0j} = -i\varepsilon_{ijk}\sigma_{1k} - \delta_{ij}\sigma_{10} = -\varepsilon_{ijk}\mathbf{U}_k + i\delta_{ij}\mathbf{\Theta}_1$$
$$\mathbf{\Gamma}_j\mathbf{U}_i = -\sigma_{0j}\sigma_{1i} = i\varepsilon_{ijk}\sigma_{1k} - \delta_{ij}\sigma_{10} = \varepsilon_{ijk}\mathbf{U}_k + i\delta_{ij}\mathbf{\Theta}_1$$
$$\mathbf{V}_i\mathbf{\Gamma}_j = -\varepsilon_{ijk}\mathbf{V}_k + i\delta_{ij}\mathbf{\Theta}_2$$
$$\mathbf{\Gamma}_j\mathbf{V}_i = \varepsilon_{ijk}\mathbf{V}_k + i\delta_{ij}\mathbf{\Theta}_2 \quad (4B)$$
$$\mathbf{W}_i\mathbf{\Gamma}_j = -\varepsilon_{ijk}\mathbf{W}_k + i\delta_{ij}\mathbf{\Theta}_3$$
$$\mathbf{\Gamma}_j\mathbf{W}_i = \varepsilon_{ijk}\mathbf{W}_k + i\delta_{ij}\mathbf{\Theta}_3$$

and

$$\mathbf{\Gamma}_i\mathbf{\Gamma}_j = -\varepsilon_{ijk}\mathbf{\Gamma}_k - \delta_{ij}\mathbf{I}, \quad \mathbf{\Theta}_i\mathbf{\Theta}_j = -\varepsilon_{ijk}\mathbf{\Theta}_k - \delta_{ij}\mathbf{I}$$

$$\mathbf{\Theta}_1\mathbf{\Gamma}_k = \mathbf{\Gamma}_k\mathbf{\Theta}_1 = i\mathbf{U}_k, \quad \mathbf{\Theta}_2\mathbf{\Gamma}_k = \mathbf{\Gamma}_k\mathbf{\Theta}_2 = i\mathbf{V}_k, \quad \mathbf{\Theta}_3\mathbf{\Gamma}_k = \mathbf{\Gamma}_k\mathbf{\Theta}_3 = i\mathbf{W}_k$$

$$\mathbf{U}_k\mathbf{\Theta}_1 = \mathbf{\Theta}_1\mathbf{U}_k = i\mathbf{\Gamma}_k, \quad \mathbf{V}_k\mathbf{\Theta}_2 = \mathbf{\Theta}_2\mathbf{V}_k = i\mathbf{\Gamma}_k, \quad \mathbf{W}_k\mathbf{\Theta}_3 = \mathbf{\Theta}_3\mathbf{W}_k = i\mathbf{\Gamma}_k$$
$$\mathbf{U}_k\mathbf{\Theta}_2 = -\mathbf{\Theta}_2\mathbf{U}_k = -\mathbf{W}_k, \quad \mathbf{U}_k\mathbf{\Theta}_3 = -\mathbf{\Theta}_3\mathbf{U}_k = \mathbf{V}_k \quad (4C)$$
$$\mathbf{V}_k\mathbf{\Theta}_1 = -\mathbf{\Theta}_1\mathbf{V}_k = \mathbf{W}_k, \quad \mathbf{V}_k\mathbf{\Theta}_3 = -\mathbf{\Theta}_3\mathbf{V}_k = -\mathbf{U}_k$$
$$\mathbf{W}_k\mathbf{\Theta}_1 = -\mathbf{\Theta}_1\mathbf{W}_k - \mathbf{V}_k, \quad \mathbf{W}_k\mathbf{\Theta}_2 = -\mathbf{\Theta}_2\mathbf{W}_k = \mathbf{U}_k.$$

This 16-element group contains three 8-element subgroups, namely, $\{\mathbf{I}, \mathbf{\Gamma}_1, \mathbf{\Gamma}_2, \mathbf{\Gamma}_3, \mathbf{\Theta}_1, \mathbf{U}_1, \mathbf{U}_2, \mathbf{U}_3\}$, $\{\mathbf{I}, \mathbf{\Gamma}_1, \mathbf{\Gamma}_2, \mathbf{\Gamma}_3, \mathbf{\Theta}_2, \mathbf{V}_1, \mathbf{V}_2, \mathbf{V}_3\}$ and $\{\mathbf{I}, \mathbf{\Gamma}_1, \mathbf{\Gamma}_2, \mathbf{\Gamma}_3, \mathbf{\Theta}_3, \mathbf{W}_1, \mathbf{W}_2, \mathbf{W}_3\}$. These three subgroups with eight elements of 4x4 matrices satisfy the closure and associative properties. They differ from the non-associative octonions algebra which will be discussed in details later in Sec. 2.3.

Here, we examine the mass-energy relation based on the operators according to the first sub-group $\{\mathbf{I}, \mathbf{\Gamma}_1, \mathbf{\Gamma}_2, \mathbf{\Gamma}_3, \mathbf{\Theta}_1, \mathbf{U}_1, \mathbf{U}_2, \mathbf{U}_3\}$. We can extend Dirac's equation involving four gamma matrices to higher dimensions involving eight direct-product operators as

$$m_0 = iE\mathbf{\Theta}_1 + \sum_{k=1}^{3} P_k\mathbf{\Gamma}_k + \sum_{k=1}^{3} Q_k\mathbf{U}_k \quad (5A)$$

By taking the square of both sides of the equation, one obtains

$$m_0^2 = E^2 + \sum_{k=1}^{3} P_k^2 \mathbf{\Gamma}_k^2 + \sum_{k=1}^{3} Q_k^2 \mathbf{U}_k^2 + \sum_{i \neq j=1}^{3} P_i P_j \{\mathbf{\Gamma}_i, \mathbf{\Gamma}_j\} + \sum_{i \neq j=1}^{3} Q_i Q_j \{\mathbf{U}_i, \mathbf{U}_j\}$$
$$+ \sum_{i,j=1}^{3} P_i Q_j \{\mathbf{U}_i, \mathbf{\Gamma}_j\} + iE \sum_{k=1}^{3} (P_k \{\mathbf{\Theta}_1, \mathbf{\Gamma}_k\} + Q_k \{\mathbf{\Theta}_1, \mathbf{U}_k\}) \quad (5B)$$

or, equivalently, according to the multiplication rules in Fig. 1, one has

$$E^2 = m_0^2 + \sum_{k=1}^{3}\left(P_k^2 + Q_k^2\right)^2 - i\Theta_1 \sum_{k=1}^{3} P_k Q_k - iE \sum_{k=1}^{3}(P_k U_k + Q_k \Gamma_k) \tag{5C}$$

Similarly, for the second assignments using $\{\mathbf{I}, \mathbf{\Gamma}_1, \mathbf{\Gamma}_2, \mathbf{\Gamma}_3, \mathbf{\Theta}_2, \mathbf{V}_1, \mathbf{V}_2, \mathbf{V}_3\}$, one has

$$\begin{aligned}m_0 &= iE\Theta_2 + \sum_{k=1}^{3} P_k \Gamma_k, + \sum_{k=1}^{3} Q_k V_k, \\ E^2 &= m_0^2 + \sum_{k=1}^{3}\left(P_k^2 + Q_k^2\right)^2 - i\Theta_2 \sum_{k=1}^{3} P_k Q_k - iE \sum_{k=1}^{3}(P_k V_k + Q_k \Gamma_k).\end{aligned} \tag{5D}$$

and for $\{\mathbf{I}, \mathbf{\Gamma}_1, \mathbf{\Gamma}_2, \mathbf{\Gamma}_3, \mathbf{\Theta}_3, \mathbf{W}_1, \mathbf{W}_2, \mathbf{W}_3\}$, one obtains

$$\begin{aligned}m_0 &= iE\Theta_3 + \sum_{k=1}^{3} P_k \Gamma_k, + \sum_{k=1}^{3} Q_k W_k, \\ E^2 &= m_0^2 + \sum_{k=1}^{3}\left(P_k^2 + Q_k^2\right)^2 - i\Theta_3 \sum_{k=1}^{3} P_k Q_k - iE \sum_{k=1}^{3}(P_k W_k + Q_k \Gamma_k).\end{aligned} \tag{5E}$$

In Equations (5A), (5D) and (5E), the last two terms of the equations involve operators $\Theta_k, \Gamma_k, U_k, V_k$ or $W_k$ which lead to mass-energy oscillations in time for a lepton and quark. Therefore, such results are neither consistent with experimental observations nor in agreement with Einstein's mass-energy relation of $E^2 = m_0^2 + \sum_k\left(P_k^2 + Q_k^2\right)$ which contains an additional kinetic energy $\sum_k Q_k^2$ due to a particle's internal structural dynamics. Therefore, although the 16-element group of direct-product operator contains three 8-element subgroup, the above direct-product model cannot represent three generations of leptons and quarks. In the following sections, we will discuss the non-associative sedenion and octonion algebras, and show that they do not encounter these problems facing the associative direct-product operator model.

**2.3. Non-associative octonion algebra and one fermion generation**

We consider the octonion algebra to avoid the problems facing the direct-product matrix model. Any element $x$ and its conjugate $\bar{x}$ in the octonion algebra, $x$ can be expressed in terms of the identity operator $e_0$ and seven other octonion unit operators

$$x = x_0 e_0 + X, \quad \bar{x} = x_0 e_0 - X, \quad X = \sum_{k=1}^{7} x_k e_k, \tag{6A}$$

where $e_k$ satisfies the anti-commutative relation $\{e_i, e_j\} = 0$, $i \neq j$ for a different pair of indices. These non-associative octonion operators follow the specific multiplication rules in Fig. 2.

| | $e_0$ | $e_1$ | $e_2$ | $e_3$ | $e_4$ | $e_5$ | $e_6$ | $e_7$ |
|---|---|---|---|---|---|---|---|---|
| $e_1$ | | $-e_0$ | $e_3$ | $-e_2$ | $e_5$ | $-e_4$ | $-e_7$ | $e_6$ |
| $e_2$ | | $-e_3$ | $-e_0$ | $e_1$ | $e_6$ | $e_7$ | $-e_4$ | $-e_5$ |
| $e_3$ | | $e_2$ | $-e_1$ | $-e_0$ | $e_7$ | $-e_6$ | $e_5$ | $-e_4$ |
| $e_4$ | | $-e_5$ | $-e_6$ | $-e_7$ | $-e_0$ | $e_1$ | $e_2$ | $e_3$ |
| $e_5$ | | $e_4$ | $-e_7$ | $e_6$ | $-e_1$ | $-e_0$ | $-e_3$ | $e_2$ |
| $e_6$ | | $e_7$ | $e_4$ | $-e_5$ | $-e_2$ | $e_3$ | $-e_0$ | $-e_1$ |
| $e_7$ | | $-e_6$ | $e_5$ | $e_4$ | $-e_3$ | $-e_2$ | $e_1$ | $-e_0$ |

Fig. 2. The color-coded multiplication table for the eight basis octonions. According to the multiplication rules, other than the identity operator, all other seven operators anti-commute with each other, however, the multiplications are not associative. The arrays are color-coded to illustrate domains of cyclic multiplication rules for quaternions. Each color-coded domain of quaternion triplets follows cyclic multiplication rules.

For the octonion algebra of Fig. 2, the 4[th] element anti-commutes with all other operators except the identity operator. However, in Fig. 1 for the direct-product operators, the 4[th] element $a_4$, $a_8$ or $a_{12}$ is commutative with all other operators, except the identity element. This property is important for the corresponding mass-energy relation to be consistent with Einstein's relativity. Therefore, the model with seven non-associative but anti-commutative octonion operators is the correct model to describe one generation of lepton or quark. The octonion model invokes three extra degrees of freedom to represent the internal structural dynamics of a fermion as three momentum components with respect to the center-of-mass reference frame, whereas the other three anti-commutative operators represent the external degrees of freedom as three momentum operators for the particle with respect to the laboratory frame.

### 2.4. Three octonion sub-algebras in sedenion algebra and three generations of charged/neutral leptons

In the last section, we explain that the octonion algebra leads to only one fermion generation. To accommodate three generations, one needs to consider higher-dimensional hypercomplex algebra, namely, the sedenion algebra. In the sedenion algebra, it consists of 16 basis ssedenion operators $\{e_k, k=0,1,2,...,15\}$, denoted as $\{I, \Gamma_1, \Gamma_2, \Gamma_3, \Theta_1, U_1, U_2, U3, \Theta_2, V_1, V_2, V_3,, \Theta_3, W_1, W_2, W_3\}$ sequentially.

The multiplication rules for 16 sedenion basis operators are given in Fig. 3, and is different from the table for the direct-product operator model shown earlier in Fig. 31

Fig. 3. The color-coded multiplication table for 16 basis sedenion operators $\{e_k, k=0,1,2,...,15\}$. The region algebra contains three distinct types of octonion sub-algebras, denoted by $\{I, \Gamma_1, \Gamma_2, \Gamma_3, \Theta_1, U_1, U_2, U_3\}, \{I, \Gamma_1, \Gamma_2, \Gamma_3, \Theta_2, V_1, V_2, V_3\}$ and $\{I, \Gamma_1, \Gamma_2, \Gamma_3, \Theta_3, W_1, W_2, W_3\}$. Each color-coded domain containing quaternion triplets follows cyclic multiplication rules. Unlike the table in Fig. 1 for the associative direct-product matrices, the 4-th, 8-th and 12-th operators in the non-associative sedenion algebra anti-commutes with **U, V, W** operators.

The multiplication table in Fig. 3 can be rearranged in order to become Fig. 4 to illustrate that the sedenion algebra contains five distinct quaternion algebras with each quaternion algebra representing a distinct type of SU(2) spinors with a cyclic multiplication rule.

Fig. 4. The rearranged multiplication table which is color-coded to illustrate five distinct types of SU(2) spinors, **U, V, W, Γ, Θ,** wich each type following a cyclic multiplication rule. Each type of spinor triad together with an unit operator forms a quaternion algebra.

As can be seen from Fig. 3, the sedenion algebra contains three distinct types of octonion algebra, which are denoted by $\{\mathbf{I}, \mathbf{\Gamma}_1, \mathbf{\Gamma}_2, \mathbf{\Gamma}_3, \mathbf{\Theta}_1, \mathbf{U}_1, \mathbf{U}_2, \mathbf{U}3\}$, $\{\mathbf{I}, \mathbf{\Gamma}_1, \mathbf{\Gamma}_2, \mathbf{\Gamma}_3, \mathbf{\Theta}_2, \mathbf{V}_1, \mathbf{V}_2, \mathbf{V}_3\}$ and $\{\mathbf{I}, \mathbf{\Gamma}_1, \mathbf{\Gamma}_2, \mathbf{\Gamma}_3, \mathbf{\Theta}_3, \mathbf{W}_1, \mathbf{W}_2, \mathbf{W}_3\}$.

The multiplication tables for these three distinct types of octonion algebra are illustrated in Fig. 5.

Fig. 5. (A) The multiplication rules for the octonions $\{I, \Gamma_1, \Gamma_2, \Gamma_3, \Theta_1, U_1, U_2, U_3\}$. (B) The multiplication table of $\{I, \Gamma_1, \Gamma_2, \Gamma_3, \Theta_2, V_1, V_2, V_3\}$. (C) The table of $\{I, \Gamma_1, \Gamma_2, \Gamma_3, \Theta_3, W_1, W_2, W_3\}$. Each color-coded domain of quaternion triplets follows cyclic multiplication rules.

We examine here the mass-energy relation according to the octonion algebra of $\{I, \Gamma_1, \Gamma_2, \Gamma_3, \Theta_1, U_1, U_2, U_3\}$ that represents the basis octonion operators $\{e_0, e_1, e_2, e_3, e_5, e_6, e_7, e_8\}$ sequentially. We can generalize Dirac's equation and utilize

1) $m_0 I = iE\Theta_1 + \sum_{k=1}^{3} P_k \Gamma_k + \sum_{k=1}^{3} A_k U_k$

$iE\, I = -m_0 \Theta_1 + \sum_{k=1}^{3} P_k \Theta_1 \Gamma_k, + \sum_{k=1}^{3} A_k \Theta_1 U_k = -m_0 \Theta_1 - \sum_{k=1}^{3} P_k U_k + \sum_{k=1}^{3} A_k \Gamma_k$

$-E^2 = \left(-m_0 \Theta_1 + \sum_{k=1}^{3} A_k \Gamma_k - \sum_{k=1}^{3} P_k U_k, \right)^2$

$= m_0^2 \Theta_1^2 + \sum_{k=1}^{3} A_k^2 \Gamma_k^2 + P_k^2 U_k^2$   (6A)

$+ \sum_{i \neq j}^{3} A_i A_j \{\Gamma_i, \Gamma_j\} + \sum_{\substack{\sum P_i \\ i \neq j}}^{3} P_i P_j \{U_i, U_j\} - m_0 P \{\Theta_1, \Gamma_k\} + m_0 \sum_{k=1}^{3} P_k \{\Theta_1, U_k\}$

$E^2 = m_{0,eff}^2 + \sum_{k=1}^{3} P_k^2, \quad m_{0,eff}^2 = m_0^2 + \sum_{k=1}^{3} A_k^2.$

The above result is consistent with Einstein's mass-energy relation indicating the octonion model gives rise to one generation of fermion with an internal structural dynamic. Similarly, for $\{\mathbf{I}, \Gamma_1, \Gamma_2, \Gamma_3, \Theta_2, \mathbf{V}_1, \mathbf{V}_2, \mathbf{V}_3\}$ and $\{\mathbf{I}, \Gamma_1, \Gamma_2, \Gamma_3, \Theta_3, \mathbf{W}_1, \mathbf{W}_2, \mathbf{W}_3\}$, one obtains

$$2) \quad m_0 \mathbf{I} = iE\Theta_2 + \sum_{k=1}^{3} P_k \Gamma_k + \sum_{k=1}^{3} B_k \mathbf{V}_k$$

$$E^2 = m_{0,eff}^{\ 2} + \sum_{k=1}^{3} P_k^{\ 2}, \quad m_{0,eff}^{\ 2} = m_0^{\ 2} + \sum_{k=1}^{3} B_k^{\ 2},$$

(6B)

and

$$3) \quad m_0 \mathbf{I} = iE\Theta_1 + \sum_{k=1}^{3} P_k \Gamma_k + \sum_{k=1}^{3} C_k \mathbf{W}_k$$

$$E^2 = m_{0,eff}^{\ 2} + \sum_{k=1}^{3} P_k^{\ 2}, \quad m_{0,eff}^{\ 2} = m_0^{\ 2} + \sum_{k=1}^{3} C_k^{\ 2}.$$

(6C)

The above three equations reproduce Einstein's mass-energy relation, i.e., $E^2 = m_{0,eff}^{\ 2} + \sum_k P_k^{\ 2}$, $m_{0,eff}^{\ 2} \equiv m_0^{\ 2} + \sum_k Q_k^{\ 2}$, $Q_k = A_k, B_k, C_k$ for a particle with an effective rest mass $m_{0,eff}$ which contains the kinetic energy of its internal structural dynamics. Equations (6A) - (6C) represent three generations of charged leptons, namely electron, muon, and tau. For three generations of the neutral leptons, i.e., the corresponding neutrino for each generation of leptons, we use the following assignments

$$1) \quad m_0 \mathbf{I} = iE\, i\Theta_1 + \sum_{k=1}^{3} P_k \Gamma_k + \sum_{k=1}^{3} A_k (\mathbf{V}_k + i\mathbf{W}_k)$$

$$iE\, \mathbf{I} = -m_0 \Theta_1 + \sum_{k=1}^{3} P_k \mathbf{U}_k + \sum_{k=1}^{3} A_k (\mathbf{W}_k - i\mathbf{V}_k)$$

$$-E^2 = m_0^{\ 2}\Theta_1^{\ 2} + P_k^{\ 2}\mathbf{U}_k^{\ 2} + \sum_{k=1}^{3} A_k^{\ 2} (\mathbf{W}_k - i\mathbf{V}_k)^2 \quad (7A)$$

$$- m_0 \sum_{k=1}^{3} A_k \{\Theta_1, \mathbf{W}_k - i\mathbf{V}_k\} + \sum_{i,j=1}^{3} P_i \{\mathbf{U}_i, \mathbf{W}_j - i\mathbf{V}_j\}$$

$$E^2 = m_0^{\ 2} + \sum_{k=1}^{3} P_k^{\ 2}$$

2) $m_0 \mathbf{I} = iE\, i\Theta_2 + \sum_{k=1}^{3} P_k \Gamma_k + \sum_{k=1}^{3} B_k (\mathbf{W}_k + i\mathbf{U}_k)$

$iE\, \mathbf{I} = -m_0 \Theta_2 + \sum_{k=1}^{3} P_k \mathbf{V}_k + \sum_{k=1}^{3} B_k (\mathbf{U}_k - i\mathbf{W}_k)$  (7B)

$E^2 = m_0^2 + \sum_{k=1}^{3} P_k^2$

3) $m_0 \mathbf{I} = iE\, \Theta_3 + \sum_{k=1}^{3} P_k \Gamma_k + \sum_{k=1}^{3} C_k (\mathbf{U}_k + i\mathbf{V}_k)$

$iE\, \mathbf{I} = -m_0 \Theta_3 + \sum_{k=1}^{3} P_k \mathbf{W}_k + \sum_{k=1}^{3} C_k (\mathbf{V}_k - i\mathbf{U}_k)$  (7C)

$E^2 = m_0^2 + \sum_{k=1}^{3} P_k^2$.

Owing to the absence of $\sum_k Q_k^2$ in the above results for three neutrino generations, one has a vanishingly small rest mass $E^2 \approx \sum_k P_k^2$ if $m_0$ is close to zero. Unlike the cases for the charged leptons, one has $E^2 = m_{0,eff}^2 + \sum_k P_k^{2'}$, where $m_{0,eff}^2 = m_0^2 + \sum_k Q_k^2 \approx \sum_k Q_k^2$ even if $m_0$ is close to zero. According to the above sedenion model, their masses could be close to zero and are much smaller than those of their corresponding leptons. The experimental observations of flavor mixing and the mass oscillations among three generations of neutrinos are believed to be induced by the symmetry breaking mechanism.

## 2.5. Sedenion algebra and three generations of quarks

Here, we propose assignments of sedenion operators for three quark generations.

1) $m_0 \mathbf{I} = iE\, i\Theta_1 + \sum_{k=1}^{3} P_k \Gamma_k + \sum_{k=1}^{3} (B_k \mathbf{V}_k + C_k \mathbf{W}_k)$

$iE\, \mathbf{I} = -m_0 \Theta_1 + \sum_{k=1}^{3} P_k \mathbf{U}_k + \sum_{k=1}^{3} (B_k \mathbf{W}_k - C_k \mathbf{V}_k)$  (8A)

$E^2 = m_{0,eff}^2 + \sum_{k=1}^{3} P_k^2,\ m_{0.eff}^2 = m_0^2 + \sum_{k=1}^{3} (B_k^2 + C_k^2)$

2) $m_0 \mathbf{I} = iE\, i\Theta_2 + \sum_{k=1}^{3} P_k \Gamma_k + \sum_{k=1}^{3}(C_k \mathbf{W}_k + A_k \mathbf{U}_k)$

$iE\, \mathbf{I} = -m_0 \Theta_2 + \sum_{k=1}^{3} P_k \mathbf{U}_k + \sum_{k=1}^{3}(C_k \mathbf{U}_k - A_k \mathbf{W}_k)$ (8B)

$E^2 = m_{0,eff}^2 + \sum_{k=1}^{3} P_k^2, \quad m_{0.eff}^2 = m_0^2 + \sum_{k=1}^{3}(A_k^2 + C_k^2)$

3) $im_0 \mathbf{I} = iE\, i\Theta_3 + \sum_{k=1}^{3} P_k \Gamma_k + \sum_{k=1}^{3}(A_k \mathbf{U}_k + B_k \mathbf{V}_k)$

$iE\, \mathbf{I} = -m_0 \Theta_3 + \sum_{k=1}^{3} P_k \mathbf{U}_k + \sum_{k=1}^{3}(-A_k \mathbf{V}_k + B_k \mathbf{U}_k)$ (8C)

$E^2 = m_{0,eff}^2 + \sum_{k=1}^{3} P_k^2, \quad m_{0.eff}^2 = m_0^2 + \sum_{k=1}^{3}(A_k^2 + B_k^2).$

One could also make different assignments to the generalized energy and momentum operators for other three heavier quark generations.

1) $m_0 \mathbf{I} = iE\Theta_1 + \sum_{k=1}^{3} P_k \Gamma_k + \sum_{k=1}^{3}(A_k \mathbf{U}_k + B_k \mathbf{V}_k + B_k \mathbf{W}_k)$

$iE\mathbf{I} = im_0 \Theta_1 + \sum_{k=1}^{3} P_k \mathbf{U}_k + \sum_{k=1}^{3}(A_k \Gamma_k + B_k \mathbf{W}_k - B_k \mathbf{V}_k)$ (9A)

$E^2 = m_{0,eff}^2 + \sum_{k=1}^{3} P_k^2, \quad m_{0.eff}^2 = m_0^2 + \sum_{k=1}^{3}(A_k^2 + 2B_k^2),$

2) $m_0 \mathbf{I} = iE\Theta_2 + \sum_{k=1}^{3} P_k \Gamma_k + \sum_{k=1}^{3}(C_k \mathbf{U}_k + B_k \mathbf{V}_k + A_k \mathbf{W}_k)$

$E^2 = m_{0,eff}^2 + \sum_{k=1}^{3} P_k^2, \quad m_{0.eff}^2 = m_0^2 + \sum_{k=1}^{3}(B_k^2 + 2C_k^2),$ (9B)

3) $m_0 \mathbf{I} = iE\Theta_3 + \sum_{k=1}^{3} P_k \Gamma_k + \sum_{k=1}^{3}(A_k \mathbf{U}_k + A_k \mathbf{V}_k + C_k \mathbf{W}_k)$

$E^2 = m_{0,eff}^2 + \sum_{k=1}^{3} P_k^2, \quad m_{0.eff}^2 = m_0^2 + \sum_{k=1}^{3}(C_k^2 + 2A_k^2).$ (9C)

**2.6. Mapping octonions to SU(3) generators and sedenion to SU(5) generators**

In this section, we discuss the mapping of the octonion operators to 8 SU(3) generators and the mapping of sedenion operators to 24 SU(5) generators. For each type of octonion algebra, one can show its isomorphism to Clifford algebra $C\ell(6)$ [20]. Here, we define three pairs of fermion creation and annihilation operators $\alpha_k^+$ and $\alpha_k$ which satisfy the anti-commutation relations as

$$\alpha_1 = (-e_6 + ie_5)/2, \quad \alpha_2 = (-e_3 + ie_1)/2, \quad \alpha_3 = (-e_7 + ie_2)/2$$
$$\{\alpha_i, \alpha_j\} = \{\alpha_i^+, \alpha_j^+\} = 0, \quad \{\alpha_i, \alpha_j^+\} = \delta_{ij}$$
(10A)

For the first lepton/quark generation, these eight basis octonion operators are denoted by $\{I, \Gamma_1, \Gamma_2, \Gamma_3, \Theta_1, U_1, U_2, U_3\}$

$$\alpha_1 = (-U_2 + iU_1)/2, \alpha_2 = (-\Gamma_3 + i\Gamma_1)/2, \alpha_3 = (-U_3 + i\Gamma_2)/2$$
$$a_1^+ = (U_2 + iU_1)/2, a_2^+ = (\Gamma_3 + i\Gamma_1)/2, a_3^+ = (U_3 + i\Gamma_2)/2.$$
(10B)

One can define $|i\rangle\langle j| \equiv \alpha_i^+ \alpha_j$ to construct the following eight SU(3) generators which are related to Gell-Mann's lambda matrices $\Lambda_k$ as

$$\Lambda_1 = |2\rangle\langle 1| + |1\rangle\langle 2| = i(U_3 - U_2)/2$$
$$\Lambda_2 = -i|1\rangle\langle 2| + i|2\rangle\langle 1| = -i(U_1 - \Theta_1)/2$$
$$\Lambda_3 = |1\rangle\langle 1| - |2\rangle\langle 2| = i(\Gamma_3 - \Gamma_2)/2$$
$$\Lambda_4 = |1\rangle\langle 3| + |3\rangle\langle 1| = i(\Theta_1 - \Gamma_2)/2$$
$$\Lambda_5 = -i|1\rangle\langle 3| + i|3\rangle\langle 1| = --i(\Gamma_1 + U_3)/2$$
(10C)
$$\Lambda_6 = |2\rangle\langle 3| + |3\rangle\langle 2| = -i(\Gamma_1 + U_2)/4$$
$$\Lambda_7 = -i|2\rangle\langle 3| + i|3\rangle\langle 2| = -i(\Theta_1 - \Gamma_3)/2$$
$$\Lambda_8 = (|1\rangle\langle 1| + |2\rangle\langle 2| - 2|3\rangle\langle 3|)/\sqrt{3} = i(\Gamma_3 + \Gamma_2 - 2U_2)/2.$$

We have shown above the eight lambda matrices as the SU(3) generators which can be constructed from $\{I, \Gamma_1, \Gamma_2, \Gamma_3, \Theta_1, U_1, U_2, U_3\}$. Because the sedenion algebra is shown to consist of three octonion sub-algebras. Similarly, one can use two other octonion basis sets, i.e., $\{I, \Gamma_1, \Gamma_2, \Gamma_3, \Theta_2, V_1, V_2, V_3\}$ and $\{I, \Gamma_1, \Gamma_2, \Gamma_3, \Theta_3, W_1, W_2, W_3\}$ to construct two other lambda-matrix generators. Together with the U-type, V-type and W-type octonion algebras which are the three distinct sub-algebra of the sedenion algebra, we can construct all together 24 generators for SU(5).

By extending the three pairs of creation and annihilation and creation operators for the octonion algebra, we define five pairs of fermion creation and annihilation operators for the sedenion algebra as

$$\alpha_1 = (-e_6 + ie_5)/2, \quad \alpha_2 = (-e_3 + ie_1)/2, \quad \alpha_3 = (-e_7 + ie_2)/2$$
$$\alpha_5 = (-e_{14} + ie_{13})/2, \quad \alpha_6 = (-e_{11} + ie_9)/2 \quad (11A)$$
$$\{\alpha_i, \alpha_j\} = \{\alpha_i^+, \alpha_j^+\} = 0, \quad \{\alpha_i, \alpha_j^+\} = \delta_{ij},$$

or, equivalently

$$\alpha_1 = (-\mathbf{U}_2 + i\mathbf{U}_1)/2, \quad \alpha_2 = (-\mathbf{\Gamma}_3 + i\mathbf{\Gamma}_1)/2, \quad \alpha_3 = (-\mathbf{U}_3 + i\mathbf{\Gamma}_2)/2$$
$$a_4 = (-\mathbf{V}_2 + i\mathbf{V}_1)/2, \quad a_5 = (-\mathbf{W}_2 + i\mathbf{W}_1)/2. \quad (11B)$$

Similar to Eq. (10C) by constructing eight of 3x3 lambda matrices for the SU(3) generators from three pairs of creation and annihilation operators, one can construct from five pairs of fermion[c creation and annihilation operators, a total of twenty-four generalized lambda matrices as the SU(5) generators. Using the five pairs of fermionic creation and annihilation operators, one can construct 10 pairs of 5x5 off-diagonal SU(5) generator matrices as $|m\rangle\langle n| + |n\rangle\langle m|$, $-i|m\rangle\langle n| + i|n\rangle\langle m|$, for $m \neq n = 1,2,...,5$. And, with the same indices one can construct four off-diagonal but orthogonal 5x5 matrices. Therefore, among these twenty-four SU(5) generators which can be represented by 5x5 generalized lambda matrices, there are four diagonal matrices and twenty off-diagonal matrices. The SU(5) symmetry plays an essential role in the GUT (grand unification theory) that has been advocated to unify the electromagnetic, weak and strong interactions among elementary particles. These three types of the 8-element octonion algebra are not totally independent of each other because they contain the same quaternion algebra $\{\mathbf{I}, \mathbf{\Gamma}_1, \mathbf{\Gamma}_2, \mathbf{\Gamma}_3\}$. Such an intersection among three octonion sub-algebras underlies the quark confinement mechanism. The cooling after the big bang of the universe plays an important role in symmetry breaking of SU(5) to become SU(3) $\otimes$ SU(2) $\otimes$ U(1), and the break sown of the sedenion algebra into direct-product of octonion and quaternion algebras. Such breakdown results in flavor mixing of neutrinos and their mass oscillations.

## 3. Conclusions

In this work, we present a theoretical analysis of higher dimensional models beyond the conventional Dirac model for a point-like fermion without a volume or an internal structure. Dirac utilized four anti-commutative matrices to couple to the first derivative of particle's space and time coordinates. Such an extension is necessary, because the traditional Dirac theory cannot explain the origins for the observed three generations of leptons and quarks postulated in the Standard Model. We first consider an 16-element group of 4x4 matrices as direct products of quaternion operators, i.e., three 2x2 Pauli's matrices and an identity matrix. This model contains three extra degrees of freedom to incorporate a particle's internal structure. This 16-element group contains three 8-element subgroups, possessing a closure property for multiplications among eight elements of each subgroup. However, this direct-product matrix model with associative operators could not reproduce Einstein's mass-energy relation, in contrast to the model with 16 sedenion basis operators. The significant difference between the two models is the associativity of whether multiplication is associative or not, as clearly illustrated from the differences in the multiplication tables of Figs. 1 and 2. The 8-element octonion algebra is shown to be able to lead to one lepton generation, however, we need 16-element sedenion algebra to encompass precisely three generation of leptons and quarks. We have shown from the color-code table arrays in Fig. 2, the sedenion algebra contains three distinct octonion algebras, with the U, V, and W-type operators, and each type of octonion sub-algebra corresponding to one generation. We have also shown for neutrinos that from Eq. (7) the effective rest mass could be vanishingly small in comparison to that in Eq. (6) for the counterpart charged leptons. We have also provided operator assignments in Equations (8) and (9) to represent three generations of lighter and heavier quarks. Moreover, we have shown that by properly pairing up the octonion basis operators one an construct 8 3x3 Gell-Mann's lambda matrices as the 8 generators for SU(3). For the U, V and W-type octonion sub-algebra of the sedenion algebra, we could pair up the sedenion basis operators to construct a total of 24 5x5 matrices to represent the 24 generators of SU(5). The SU(5) symmetry plays an essential role in the GUT that unifies the electroweak and strong interactions among the Standard Model's elementary particles. These three types of the 8-element octonion algebra contain the same quaternions $\{\mathbf{I}, \mathbf{\Gamma}_1, \mathbf{\Gamma}_2, \mathbf{\Gamma}_3\}$, and such an intersection underlies the quark confinement mechanism. Soon after the big bang of the universe, its cooling plays an important role in symmetry breaking for SU(5) to be broken down to become $SU(3) \otimes SU(2) \otimes U(1)$ [25, 26], and the broken sedenion algebra into direct-product algebra of octonions and quaternions results in flavor mixing of neutrinos to induce the mass oscillation behavior.

**Acknowledgment**

This research was funded by the Scientific Research Foundation for High-level Talents of Anhui University of Science and Technology (grant number (2022yjrc67).